# Atmospheric changes from solar eclipses

**K. L. Aplin[1], C. J. Scott[2] and S. L. Gray[2]**

*1. Department of Physics, University of Oxford, Denys Wilkinson Building, Keble Road, Oxford OX1 3RH*
*2. Department of Meteorology, University of Reading, PO Box 243, Earley Gate, Reading RG6 6BB*

# Summary

This article reviews atmospheric changes associated with 44 solar eclipses, beginning with the first quantitative results available, from 1834 (earlier qualitative, accounts also exist). Eclipse meteorology attracted relatively few publications until the total solar eclipse of 16 February 1980, with the 11 August 1999 eclipse producing the most papers. Eclipses passing over populated areas such as Europe, China and India now regularly attract scientific attention, whereas atmospheric measurements of eclipses at remote locations remain rare.  Many measurements and models have been used to exploit the uniquely predictable solar forcing provided by an eclipse. In this paper we compile the available publications and review a sub-set of them chosen on the basis of importance and novelty. Beyond the obvious reduction in incoming solar radiation, atmospheric cooling from eclipses can induce dynamical changes. Observations and meteorological modelling provide evidence for the generation of a local eclipse circulation which may be the origin of the "eclipse wind". Gravity waves set up by the eclipse can, in principle, be detected as atmospheric pressure fluctuations, though theoretical predictions are limited, and many of the data are inconclusive. Eclipse events providing important early insights into the ionisation of the upper atmosphere are also briefly reviewed.

# Main Text

1. Introduction

The weather during solar eclipses has always been of interest, because of the effect it has on the viewing experience. Early observations were compiled by Halley (1714) to deduce the path of a total solar eclipse across the UK on 22 April 1713, with the help of observers across the country. Inclement weather occasionally impeded the spectacle, as Halley explained, "My worthy collegue Dr John Keill by reason of Clouds saw nothing distinctly at Oxford but the End…". As well as the excitement of darkness falling, and birdsong stopping, Halley noted other widely-observed meteorological effects, particularly, "the Chill and Damp which attended the Darkness of this Eclipse".

The first known quantitative eclipse weather observations, on 30 November 1834, were reported in the *Boston Medical and Surgical Journal* (Anon, 1834).  Although not named, the author was a diligent observer, finding that temperature measured in the shade and, "of course, in a northern exposure", dropped by 1.5 °F from the start of the eclipse to greatest obscuration, with no substantive change in the pressure. Soon afterwards, careful qualitative observations from a UK partial eclipse of 15 May 1836 were made by Birt (1836) who used Luke Howard's then-novel cloud classifications to give a detailed account of cloud changes during the eclipse. Cumulus ("fair-weather") clouds died away as the eclipse developed, which might now be explained in terms of diminished convection from reduced solar heating.

Systematic eclipse weather campaigns were pioneered by Winslow Upton who was professor of astronomy at Brown University (Archibald, 1914). He led a series of expeditions, with his first eclipse meteorology publication from the Russian solar eclipse of 19 August 1887 (Archibald, 1914) followed by other eclipses over the USA (Upton and Rotch, 1887; 1893). His work is characterised by multi-instrumented sites with several observers focusing on different aspects to maximise the scientific return (automated records were not used). Soon after this, Clayton (1901a) provided a modernising influence, being the first to apply meteorological techniques such as use of "anomalies" i.e. subtracting the average behaviour to identify the changes caused by the eclipse. Clayton was also the first to recognise the special circumstances provided by an eclipse in imposing a well-understood diminution of solar radiation. As Clayton put it, "the eclipse may be compared to an experiment by Nature … by eliminating the influence of other known phenomena", which he used to good effect by investigating the dynamical effects of eclipses, discussed in more detail in section 2.1.

By the late twentieth century improvements in computing and atmospheric physics led to the extensive use of meteorological models. The first known application of a meteorological model to an eclipse was published by Gross and Hense (1999) in anticipation of the 11 August 1999 European total solar eclipse. Since then, there have been several sophisticated model studies of eclipse meteorology, which will be covered in section 2.2.

The final effect of eclipses on the weather to attract significant scientific attention is the generation of gravity waves, fluctuations in atmospheric pressure generated by the atmospheric cooling from the obscuration of the Sun by the Moon. This is a relatively recent concept, first predicted by Chimonas (1970) and Chimonas and Hines (1970). Anderson and Keefer (1973) planned and carried out pressure measurements on the basis of these predictions, and retrospectively analysed older eclipse pressure data for the possible existence of gravity waves. Attempts to measure gravity waves at subsequent eclipses have been relatively common, but most of the findings have been inconclusive and inconsistent, section 2.3.

* (karen.aplin@physics.ox.ac.uk)

### 1.1 Summary of previous work and structure of the paper

Atmospheric measurements during 44 eclipses are summarised in Table 1. Many articles exist describing eclipse meteorology, about 120 in total, Figure 1. Publications have become significantly more numerous since the 16 Feb 1980 eclipse, with the 11 August 1999 European total solar eclipse provoking the most reports. The number of papers written about each eclipse in recent years appears to be related to the eclipse's path – if it is over densely populated areas such as Europe, North America, India and China, the Moon's shadow will pass over existing meteorological stations or provoke atmospheric field campaigns. The role of the amateur observer can also be significant, and is discussed by Hanna (2000; 2016).

It is not possible to summarise all the results in Table 1 here (a subset of the data are tabulated by Kameda et al, 2009) so the approach taken is to focus on work that is of particular interest, usually the first of its kind, or a novel method. For example, meteorological measurements of eclipses over sparsely populated areas such as the poles remain of scientific interest due to the low sun angles and opportunities to investigate rarely-observed effects (e.g Kameda et al, 2009; Sjöblom, 2010), and these are discussed in section 2.4. This paper is organised by topic up to, but not including, results from the March 2015 eclipse, covering circulation changes (section 2.1), meteorological modelling (section 2.2), and gravity waves (section 2.3) in the troposphere and stratosphere. Other effects, such as atmospheric electricity, and novel approaches, like measurements aloft, are mentioned in section 2.4. In the upper atmosphere, here defined as 100 km upwards, photoionisation is dominant and therefore eclipses produce changes that are more significant than those seen in the troposphere and stratosphere. An entire paper could be written on historical work investigating the response of the upper atmosphere to eclipses, but we have provided a brief summary for completeness (section 3).

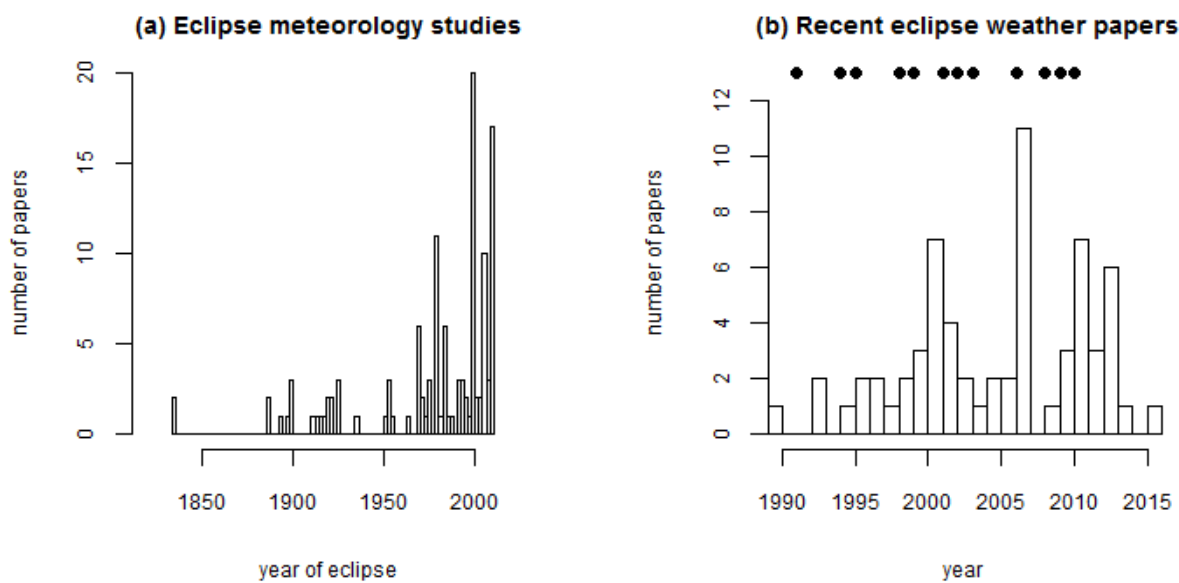

*Figure 1 Number of papers published about eclipse meteorology by year of publication (a) since the first known quantitative measurements (1834) to the present and (b) from 1990 to the present (excluding the 2015 eclipse), with the years in which eclipses occurred indicated as large black dots at the top. (Note that there were two solar eclipses in 1994 and 2003).*

2. Meteorological effects of eclipses in the troposphere and stratosphere

The obvious effect of an eclipse is that the solar radiation reaching the earth decreases as the Moon obscures the Sun's disk. In a total solar eclipse, the solar radiation drops to zero; in a partial eclipse it does not reach zero. (There is negligible difference between the atmospheric effects of partial and annular eclipses of similar magnitude, as an annular eclipse is when the Moon covers only the centre part of the Sun, leaving a ring of it showing, whereas during a partial eclipse the Moon leaves a crescent of the Sun showing). Cooling from solar eclipses is greater when the sun is higher in the sky since the sun's obscuration has a larger relative effect on the downwelling solar radiation, i.e. at local noon and near the midsummer solstice. For example, a substantial temperature change, -5 ˚C, was recorded after the 21 June 2001 total solar eclipse over northern Zimbabwe (15 ˚S), which took place near local noon, but close to the southern winter solstice (Figure 2). The maximum surface temperature changes in the literature are about -7 ˚C (e.g. Vogel et al, 2001). For the eclipse in Figure 2, the lag between totality and the temperature minimum is 30 minutes, similar to Vogel et al (2001) but longer than most other clear sky lags (e.g. Kameda et al (2009) quoted 6-30 minutes with an average of 12 minutes). This lag is thought to be caused by the thermal inertia of the surface layer (e.g. Aplin and Harrison, 2003) but it has also been reported to be empirically related to the global solar radiation $S_g$ at fourth contact (Kameda et al 2009). The sensitivity of the temperature change to solar radiation is 0.017±0.001 ˚C/(Wm$^{-2}$), consistent with other clear sky data (Kameda et al, 2009).

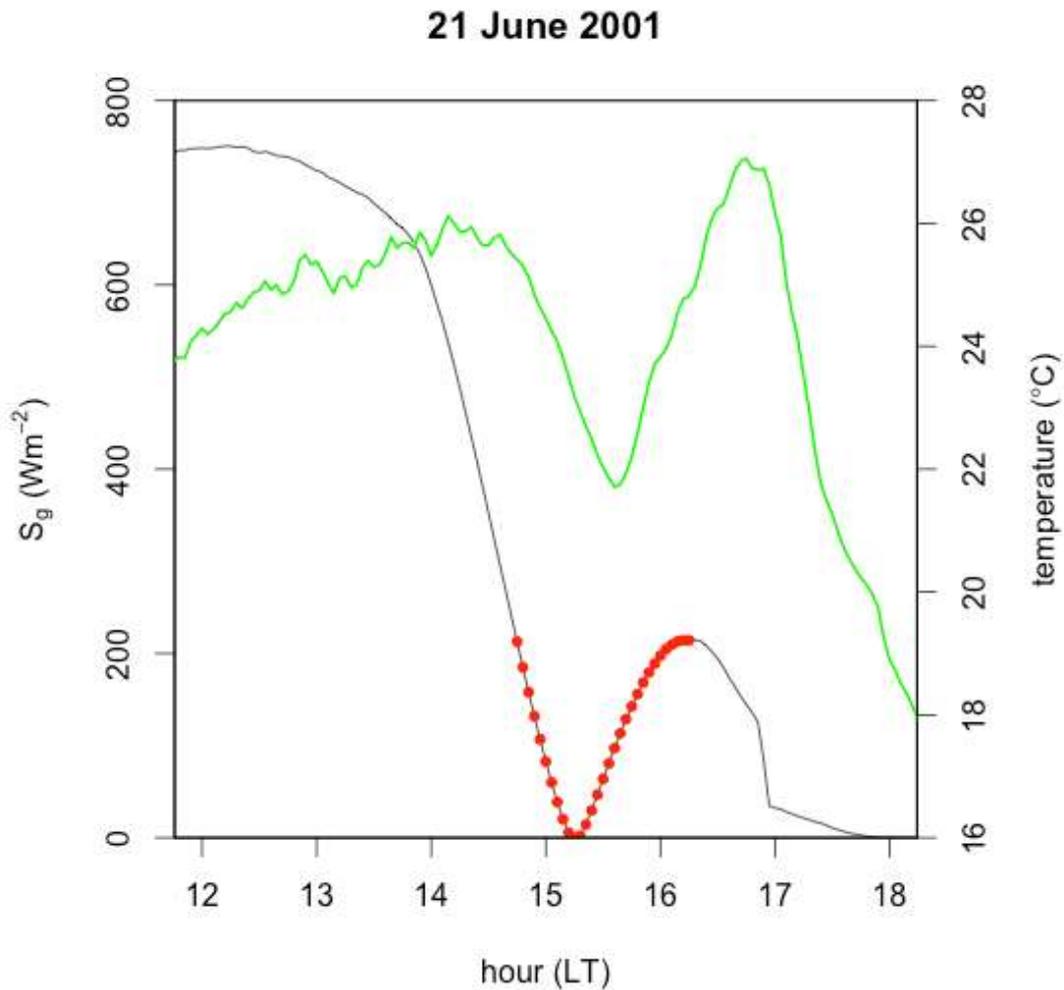

*Figure 2 Temperature change during a total solar eclipse over Mana Pools, northern Zimbabwe on 21 June 2001 (after Milford, 2001). The thin black line indicates the global solar radiation, with measurements during the eclipse period overlain as red points. The air temperature is shown as a thicker green line. All data are 5 minute averages. The time axis is in local time (LT= UT+2h). Colour version available online.*

Figure 3 is a summary of meteorological effects due to partial solar eclipses over Reading University's Atmospheric Observatory, which we believe to be the most comprehensive set of eclipse weather observations from a single site. The diminution in solar radiation is usually detectable in the global solar radiation measurements even under thick cloud, as for the two March eclipses shown, but for the 12 October 1996 eclipse no solar radiation change is evident. Both the August eclipses, which were similar in terms of time of year, day and cloud showed a clear reduction in surface temperature. The largest temperature drop was 1.5°C, on 11 August 1999, due to the 97% magnitude of the eclipse and the broken cloud cover. In comparison, the cloudy umbral region at Camborne, UK had a local temperature minimum before eclipse maximum (Aplin and Harrison, 2003).

Near-surface relative humidity changes are typically anticorrelated with near-surface air temperature changes, and increase directly as a consequence of the cooling during eclipses. For example, Namboodiri et al. (2011) reported a maximum variation of 19% with a 28-minute time lag after the maximum eclipse associated with a 3.5°C cooling during the 2010 annular eclipse over Thumba, India. A reduction of specific humidity has also been reported and attributed to eclipse-induced subsidence of drier air (Bhat and Jagannathan, 2012) but relative humidity changes are typically dominated by the temperature effect.

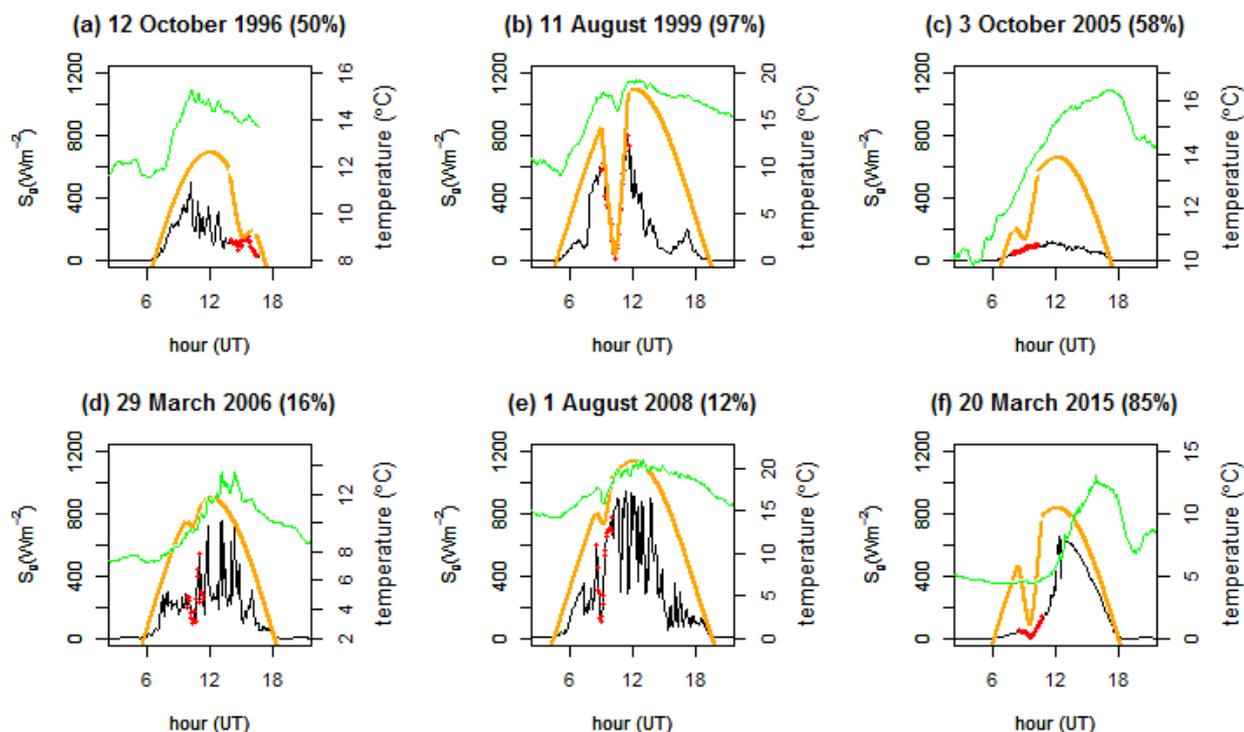

Figure 3 Meteorological effects of partial eclipses measured at Reading University Atmospheric Observatory, with the maximum obscuration (the fraction of the Sun's area covered by the Moon) indicated as a percentage (a) 12 October 1996 (50%) (no data available after 1800) (b) 11 August 1999 (97%) (c) 3 October 2005 (58%) (d) 29 March 2006 (16%) (e) 1 August 2008 (12%) and (f) 20 March 2015 (85%). The thick orange trace shows the predicted global solar radiation ($S_g$) during the eclipse using the simple model given in Aplin and Harrison (2003). The thin black line indicates the global solar radiation, with measurements during the eclipse period overlain as red points. The air temperature is shown as a thick green line. All data are 5 minute averages. Colour version available online.

Atmospheric cooling by eclipses is responsible for most of the other more subtle meteorological changes discussed below. Circulations can be modified, small pressure waves, "gravity waves" set up, and atmospheric turbulence suppressed. These can affect surface winds, temperatures and radiation, and the electrical state of the atmosphere.

2.1 Circulation changes

Eclipses often modify synoptic or mesoscale circulations in favour of their own local cooling-induced effects. Associated with these circulation changes may be an "eclipse wind" which has been reported anecdotally as darkness descends, perhaps related to the heightened emotional state of some eclipse observers (Anderson, 1999). Subrahamanyan et al (2011) found that the sea breeze circulation at Thumba, a coastal site in southern India, was suppressed during a relatively long 77% annular eclipse occurring near local noon on 10 Jan 2010. During maximum eclipse, radiosonde measurements indicate that the sea breeze cell was only about 300m thick compared to 600m under similar synoptic conditions the preceding day.

Clayton (1901a) was the first to propose that solar eclipses modified atmospheric circulation, based on a compilation of observations from several sites across the USA during the 28 May 1900 solar eclipse, as was mentioned in section 1. His results suggested a cooler anticyclonic (clockwise) circulation extending about 1500–2000 miles from the umbra, with a cyclonic (anticlockwise) circulation ring beyond, about 1000 miles across, surrounded by an outer ring of higher pressure. Modelling studies by Prenosil (2000) predicted similar effects to those of Clayton's theory for the 1999 eclipse over Europe. In high-resolution wind measurements in the umbra and penumbra of the 11 August 1999 eclipse, Aplin and Harrison (2003) found that the eclipse circulation was roughly consistent with Clayton's model for the penumbra, but that only the umbral region acted as a cold-cored cyclone, with a cold outflow into a surrounding anticyclonic region. Subsequent studies comparing observations from the Met Office surface station network with a forecast from its Unified Model (which did not represent the eclipse) found that the circulation changes seen were likely to be eclipse-induced (Gray and Harrison, 2012). These circulation changes could contribute to the sensation of an "eclipse wind" reported by some observers.

2.2 Modelling studies

There are a range of model studies of eclipses, summarised in Table 2. The two main categories are discussed here: firstly meteorological models (often from national weather services), which represent most of the recent output, in section 2.2.1 and secondly, a few solar radiation models, covered in section 2.2.2.

2.2.1   Meteorological models

One of the underlying problems of eclipse meteorology is distinguishing changes caused by eclipses from other, independent, atmospheric effects. This situation is most complicated when meteorological changes are occurring simultaneously with an eclipse, for example when a weather front is arriving at the observation site. The standard approach used is (a) to compare the data with days before and after the eclipse, or (b) to interpolate the changes during the eclipse period to estimate what would have happened if there was no eclipse. As pointed out by Gray and Harrison (2012), neither of these approaches is ideal since (a) requires consistent conditions on the days either side of the eclipse and (b) neglects temporal lags in eclipse-induced changes, which may persist after the eclipse. Under these circumstances in particular, meteorological models can play an important role, as was hinted in section 2.1 above.

Gross and Hense (1999) used the (now no longer operational) German Weather Service forecast model with 14 km grid spacing to predict the effect that the 1999 solar eclipse trajectory would have had over Central Europe for the weather conditions on 11 August 1998, a clear and hot day, and found that a near-midday eclipse six weeks away from the summer solstice would cause a substantial (3.5 ˚C on average) and rapid temperature drop, and that the atmospheric circulation would be affected. Unfortunately, as Aplin and Harrison (2003) reported, a weather front in North Western Europe meant that the weather, and observed effects, could differ substantially from clear sky values. Prenosil (2000) used a low-resolution (63.5 km grid spacing) regional weather prediction model to simulate the tropospheric response to the total solar eclipse on the 11 August 1999 over central Europe both with real time data for that day, and for a water vapour and cloud-free scenario. A cold-cored cyclone circulation was expected, and the maximum temperature change of -2 ˚C was predicted in an area north of Dover, UK and east of Calais, France; both predictions were not inconsistent with the observations by Aplin and Harrison (2003). Vogel et al. (2001) also performed numerical experiments for the 11 August 1999 eclipse, using a high resolution (4-km grid spacing), non-hydrostatic regional atmospheric model (KAMM, Adrian and Fieldler (1991)) coupled to a surface vegetation model and positioned over the upper Rhine valley. An idealised, spatially- and temporally-constant, background flow and cloud-free conditions were assumed. Temperature decreases of 6-7°C were simulated. The modification of winds on slopes in the inhomogeneous terrain led to large changes in wind direction. However, for homogeneous terrain, small effects on wind field were found, with "no implication of the so called eclipse wind". Founda et al (2007) performed numerical model experiments with the Weather Research and Forecast (WRF) model for the 29 March 2006 eclipse. They used nested domains with 2 km grid spacing in the highest-resolution domain. In this domain they predicted eclipse-induced 1.5 m temperature anomalies that were consistent in magnitude and timing with those observed (e.g. predicted cooling of 2.8˚C in the centre of Athens near eclipse maximum compared to 2.6˚C observed). Some wind speed and direction changes were measured in totality, but Founda et al (2007) attributed them to a combination of the synoptic evolution and changes in the local sea breeze flow.

Eckermann et al (2007) used a US Department of Defense model with a "high top" of 0.001hPa for the eclipse of 4 December 2002. This model, including a radiative code to predict the absorption of atmospheric species such as water vapour and ozone, was used to predict surface temperature and pressure changes by differencing runs initialised for identical pre-eclipse conditions with and without the eclipse trajectory included. Surface temperature changes were greatest over land, with a cooling of 4 ˚C, consistent with observations for other eclipses. Surface pressure fluctuations of 0.1-1hPa were also predicted from gravity waves (similar to the 0.3 hPa predicted by Prenosil (2000)) which will be discussed in the next section. A further application of meteorological modelling demonstrated for this work is that

*Phil. Trans. R. Soc. A.*

atmospheric effects due to eclipses can be predicted in remote areas, such as for the path of the 2002 eclipse over southern Africa, the southern Indian Ocean and Australia. To our knowledge, this eclipse yielded no meteorological observations due to its trajectory over sea and sparsely populated areas.

One of the problems of modelling eclipses is that a numerical weather model run with and without an eclipse gives results that represent the eclipse in the model, but do not verify the model itself. To move away from this, Gray and Harrison (2012) employed a different methodology. A high spatial and temporal resolution model was initialised with pre-eclipse weather conditions and run with no knowledge of the eclipse. This model run, when compared to observations made at a network of UK Met Office stations, and those in Aplin and Harrison (2003), could be used to verify the model's prediction of the changes caused by the eclipse over the UK. Over a sub-selected penumbral region excluding the south-western peninsula (chosen to correspond to the clearer sky, and better instrumented region) a reduction in wind speed of 0.7ms$^{-1}$ and a backing of the wind direction, i.e. an anticlockwise change corresponding to a more easterly direction, were attributed to the eclipse. This wind direction change, whilst clearly not associated with an actual gust of wind, also supports the concept of an "eclipse wind", although it is not clear which, if any, of these effects would be detectable by a human observer

### 2.2.2 Solar radiation models

In addition to standard meteorological models, eclipses provide an ideal opportunity for testing solar radiation models. These models mainly simulate the path of photons in the eclipsed region. Whilst this situation is relatively well understood for partial eclipses (e.g. Shaw, 1978), total eclipses present technical challenges, since photons detected within the umbra must have arrived by multiple scattering, and therefore three-dimensional scattering code is needed. These models have contributed to understanding of mainly small effects such as the effect of aerosols on the visible radiation and the contribution of the solar corona to the radiation (e.g. Emde and Mayer, 2007). Limb darkening, changes in the spectral irradiance across the sun's disk, can also be predicted during solar eclipses using radiative models (Koepke et al, 2001).

### 2.3 Gravity waves

Gravity waves are atmospheric pressure waves, first predicted as a consequence of the transient cooling induced by a solar eclipse by Chimonas (1970) and Chimonas and Hines (1970). Specific predictions were made for the 7 March 1970 eclipse, and, in response, Anderson et al (1972) made sensitive pressure measurements and surveyed historical data. Many of the historical pressure measurements reviewed showed fluctuations of about 24 Pa (1 Pa = 0.01 hPa) with duration of about half the eclipse. The measurements made by Anderson et al (1972) showed a periodicity of 89 min with magnitude of 25 Pa, consistent with the earlier work. These fluctuations were an order of magnitude greater than the 0.2 Pa changes predicted by Chimonas (1970). More recent predictions are larger, 10-100 Pa (Prenosil, 2000; Eckermann et al, 2007), and the range of periodicities actually detected during solar eclipses, summarised in Table 3, have periodicities of 20 min up to several hours.

The variability in both measurements and theoretical predictions originates from the range of atmospheric layers from which eclipse-induced gravity waves can be generated, and because several mechanisms are involved. In the stratosphere and troposphere, solar UV radiation is converted into heat (infra-red radiation) by ozone and water vapour respectively. When the solar UV radiation is attenuated, at night or during a solar eclipse, these layers of the atmosphere are cooled and a gravity wave is induced. The daily gravity wave is known as the atmospheric solar tide; further gravity waves are therefore expected to emanate from the penumbral and umbral regions of solar eclipses (Marty et al, 2013). Several theories exist explaining exactly how these gravity waves, are generated and the effects that can be expected from them, but limited predictions are available for surface pressure changes (Marty et al, 2013 and references therein) and the existing observations are only approximately consistent both with each other and the theoretical work. Measurements made so far are summarised graphically in Figure 4.

9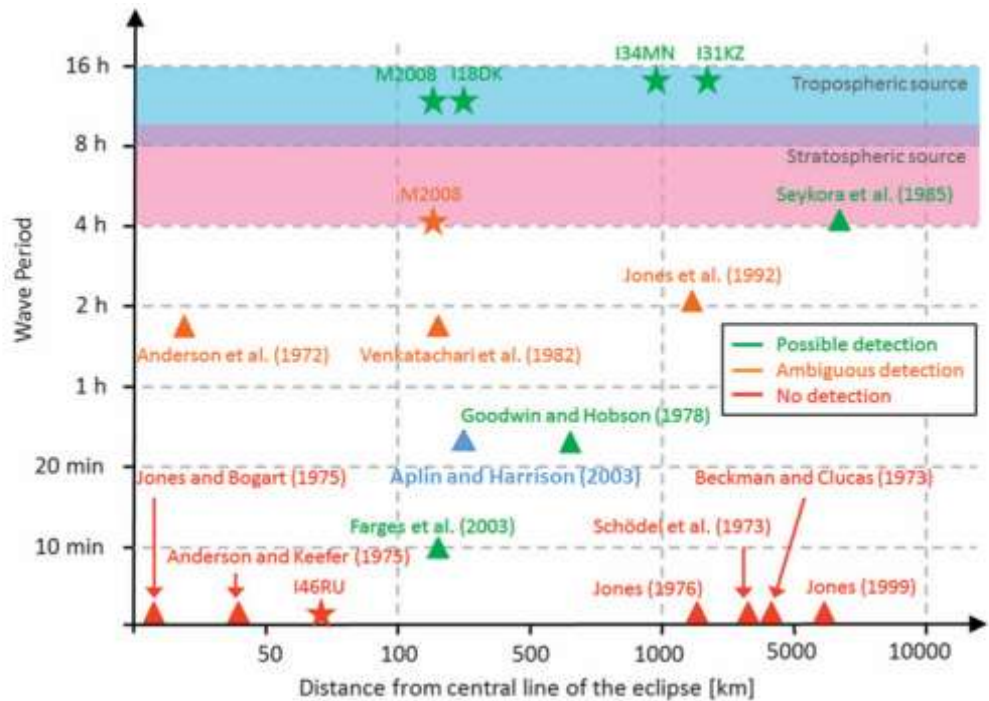

*Figure 4 Summary of eclipse gravity wave observations, adapted from Marty et al (2013). Measurements not included by Marty et al (2013) are in blue. Measurements made by Marty et al (2013) are stars and those made by others are triangles.*

## 2.4 Other measurements

In this section other eclipse weather measurements of interest from Table 1 are briefly reviewed.

### 2.4.1 Atmospheric electricity

The first measurements to be made at eclipses beyond the standard meteorological quantities of pressure, temperature, wind, humidity and solar radiation, were of atmospheric electricity. Atmospheric electricity refers to both thunder and lightning, and the subtler electrical properties of the atmosphere away from thunderstorms, such as the air's electrical conductivity. In the early years of the twentieth century atmospheric electricity was a scientific priority due to the recent discovery of radioactivity, and its ability to ionise the air. It was well known at the time that natural ionisation led to the slight electrical conductivity of air, but the origin of "penetrating radiation" which ionised air through thick layers of lead would not be fully understood until Hess's discovery of cosmic rays in 1912 (Aplin et al, 2008). Although the full text of Knoche and Laub's (1917) report on atmospheric electrical measurements during eclipses is not available, solar effects on atmospheric electricity would have clearly been of scientific interest and may have motivated early eclipse experiments. An additional source of atmospheric electrical observations was from magnetic observatories, such as the magnetic data from the 21 August 1914 total solar eclipse compiled by Bauer and Fisk (1916), which includes atmospheric electrical and meteorological measurements. Bennett (2016) gives a more detailed account of atmospheric electrical measurements during solar eclipses.

### 2.4.2 Measurements aloft

The first successful measurements away from the surface during a solar eclipse were made by Samuels (1925) who arranged for six kites, each carrying a meteograph, to fly from several sites in the penumbra of a solar eclipse over the USA on 24 January 1925. As the eclipse occurred at approximately 0800 local time, when the sun was low in the winter sky, only two of the six flights showed any detectable response. The kites were at 400-800 m during the eclipse with the largest temperature change at a North Dakota site, where the eclipse magnitude was 95%. (The earliest attempt at eclipse measurements aloft was the launch of two balloons before and after maximum obscuration from Eskdalemuir Magnetic Observatory in Scotland on 21 August 1914, but the balloons were lost, so no data were retrieved (Bauer and Fisk, 1916).)
*Phil. Trans. R. Soc. A.*

No further above-ground measurements during eclipses were attempted, as far as we know, until instrumented rocket launches in the 1970s (e.g. Quiroz and Henry, 1973). In the twenty-first century radiosonde (meteorological weather balloon) launches during eclipses have become more popular (Satyanarayana et al, 2002; Subrahamanyan et al, 2011; Harrison et al, 2016) with some limited recent rocket launches (Namboodiri et al, 2011).

### 2.4.3 Measurements near the poles

As discussed in section 2, measurements of the atmospheric effects of eclipses are most common over populated areas. The unpopulated environments of the poles receive little attention; for example, Kameda et al (2009) provide evidence that 18 total solar eclipses over Antarctica were completely unobserved since the continent was first visited by Captain Cook in the 1770s. In what were probably the first polar meteorological measurements during a total solar eclipse – and almost certainly the first human sighting of any Antarctic eclipse - Kameda et al (2009) and colleagues from the Japanese Antarctic Research Expedition observed the eclipse of 23 November 2003 from Dome Fuji (77 °S, 3810m). The temperature of –51°C dropped by 3 °C in response to the eclipse, although the maximum solar radiation in cloudless skies was only 200 Wm$^{-2}$. No clear effects on atmospheric pressure or wind were detected. Whilst the temperature change observed was typical, the time lag between the end of totality and the temperature minimum was longer for this eclipse than for others, which motivated Kameda et al (2009) to compare the time lag with minimum temperature against the clear sky global solar radiation for several eclipses, as in section 2 above.

The eclipse measurements closest to the poles were published by Sjöblom (2010) who used five sites near Longyearbyen on the Svalbard archipelago at 78 °N. The eclipse of 1 August 2008 was of particular interest because, although partial (93%) at that site, the midnight sun was obscured so the eclipse had a significant effect because of the lack of diurnal variation in solar radiation. During and after the eclipse katabatic winds started to flow down to the valleys linking to the large fjord in which Longyearbyen is located, after which stratus clouds started to form, which advected onto land as fog. This fog persisted for three days and all air traffic into and out of the archipelago was stopped. There were few explanations for the fog other than the eclipse since there were no synoptic changes, nor, of course, a sunrise or sunset to modify the local surface radiation balance. Downslope winds are known to occur in mountainous environments as a consequence of local cooling similar to that induced by an eclipse (Sjöblom, 2010). This eclipse has therefore also been the eclipse with the greatest economic impact, even though relatively few people were in the path of the Moon's shadow.

### 3. Eclipse effects on the upper atmosphere

There have been extensive studies of the Earth's ionosphere and upper atmosphere during solar eclipses, as it was recognised early in the twentieth century that an eclipse could provide a useful test of scientific theories explaining the behaviour of the upper atmosphere, here defined as from about 100km altitude upwards.

### 3.1 Scientific motivation

The ionosphere is a weakly charged plasma layer of the upper atmosphere, largely generated when atmospheric gases are ionised by incoming solar Extreme Ultraviolet and x-ray radiation and lost through recombination reactions. At any given time, the rate of change of the ion number concentration, *N*, in a given volume can be represented by the expression;

$$\frac{dN}{dt} = q - L - T \tag{1}$$

where *q* is the ion production rate, *L* is the ion loss rate and *T* is the loss rate through ion transport. We now know that, in the ionospheric E region (at around 100 km), the dominant neutral gases are molecular. When ionised, these molecular ions recombine with free electrons, and any excess electron energy causes the neutral molecule formed to break up in a process known as dissociative recombination. The average lifetime of an ion in the E region is of the order of seconds and so the transport term in equation 1 is effectively zero. At the higher altitudes of the F2 region (about 300 km) the dominant ion species is O$^+$ which, due to the lower gas densities and the atomic ions' inability to recombine with energetic electrons, has a longer average lifetime (~10 s) and transport becomes important. The intervening F1 layer therefore represents a transition region between molecular and atomic chemistry.
Before spaceflight or even sounding rockets could reach such lofty altitudes, the exact composition of the upper atmosphere remained unknown and solar eclipses were seen as an opportunity to measure the ionospheric decay rate in order to determine which species were involved.

### 3.2 Measurements

Techniques were first developed to determine the ion number density using radio (Briet and Tuve, 1926). A radio signal is returned from the ionosphere when it reaches an altitude where the local plasma frequency matches the radio frequency. Since the local plasma frequency is proportional to the square-root of the ion number density, transmitting radio pulses with a range of shortwave frequencies revealed the height profile of free electrons, with the time-of-flight of the signal being used to estimate the height of the ionospheric layers.

During a solar eclipse, with ion production removed at totality, the loss rate, $L$, was estimated from the time delay between totality and the subsequent minimum in atmospheric ionisation. These experiments invariably under-estimated the loss rate due to the fact that, while emissions from the visible solar disc were obscured by the Moon at totality, emissions from the solar corona were not. Using eclipse experiments to determine loss rates fell out of fashion after sounding rockets were developed that could make direct measurements of upper atmospheric composition. Now that these loss rates are well known, these early eclipse experiments have been re-interpreted to estimate how the relative size of the solar corona has varied throughout the 20$^{th}$ century (Davis et al, 2001, Davis et al, 2009). These two papers contain references to the original research publications, along with details of previously unpublished eclipse measurements obtained from international data archives.

Early studies were not restricted to investigations of the loss rate. Beynon and Brown (1956) list details of 22 eclipse experiments prior to 1956 and report details from a symposium held at the Royal Society in London in August 1956 in which experiments conducted during eclipses in the 1950s were discussed. Topics included studies of solar radiation, solar radio noise, ionospheric recombination, and eclipse-induced geomagnetic effects.

Modern eclipse studies have benefited from the advent of advanced radar facilities (e.g. Pitout et al, 2013; Wang Hu et al, 2011) and have concentrated on radio propagation studies (e.g. Guha et al, 2012), wave propagation (e.g. Sekar et al, 2014), and the impact of solar eclipses on the upper atmosphere as a way of verifying complex atmospheric models (e.g. Muller-Wodarg et al, 1998; Le et al, 2008; Le et al, 2009).

### 4. Conclusions

Solar eclipses offer a unique opportunity to measure the atmospheric effects of a well-defined attenuation of the incoming solar radiation, an aspect first noticed by Clayton (1901a) who combined multiple observations to make pioneering deductions about circulation changes from eclipses. Clayton's theory was not universally accepted at the time (Bigelow, 1901; Clayton, 1901b) and while more recent work has found eclipse-induced circulations consistent with Clayton's theory (Gray and Harrison, 2012; Aplin and Harrison, 2003), other work has found either no circulation changes (Eaton et al., 1997) or attributed these changes to changes in the synoptic situation or in local mesoscale flows such as sea breezes or mountain slope flows (Vogel et al., 2001; Namboodiri et al., 2011; Founda et al., 2007; KiranKumar et al. 2013). Clayton's work also represented a transition between the purely empirical and observational aspects of pre-1900 eclipse weather, to the twentieth-century approach of synthesising measurements and searching for broader atmospheric effects beyond surface changes in solar radiation and temperature. In the first half of the twentieth century solar eclipses offered a much-needed opportunity to study the upper atmosphere, which was then almost totally inaccessible, and in the second half the focus shifted to understanding the gravity waves produced by eclipses and their propagation through the atmosphere. In the late twentieth and early twenty-first centuries the use of models has become widespread in eclipse meteorology, both to predict the atmospheric changes caused by eclipses, and to present a verification opportunity for the model.

*Phil. Trans. R. Soc. A.*

Eclipse meteorology is growing as a subject with each eclipse over populated areas now attracting many publications, most notably the 1999 and 2006 eclipses over Europe. There are now over 100 scientific papers reporting highly diverse aspects of eclipse meteorology and it has not been possible to cover them all here, beyond summarising pioneering work and the significant atmospheric effects. The major open scientific question in eclipse meteorology is the origin and effect of gravity waves, which are still poorly understood with many conflicting theories and observations. The detailed circulation changes from eclipses, particularly their extent (or not) under cloudy skies, also require further investigation. It is likely that the forthcoming 2017 total solar eclipse passing from west to east over the USA will stimulate further work in this area.

# Additional Information

**Acknowledgments**
Dr J. Milford (formerly of Department of Physics, Zimbabwe University) provided the data shown in Figure 2. Prof R.G. Harrison (Department of Meteorology, University of Reading) provided the data shown in Figure 3.

Eclipse ephemera were obtained from http://www.timeanddate.com/, http://calsky.com/ and http://eclipse.gsfc.nasa.gov/eclipse.html.

**Data Accessibility**
Data is available on request from the corresponding author.

**Competing Interests**
None.

**Author Contributions**
KLA carried out the data analysis and literature search, and prepared the plots and tables. All authors contributed to manuscript drafting and review, and gave the final approval for publication.

# References

Abbott, W. N., (1958) On certain radiometric effects during the partial solar eclipse of February 25, 1952. Geofisica pura e applicata 39.1, 186-193.

Adrian, G., F. Fiedler (1991), Simulation of unstationary wind- and temperature fields over complex terrain and comparison with observations, *Contrib. Phys. Atmos.*, 64, 27–48

Amiridis, V., D. Melas, D. S. Balis, A. Papayannis, D. Founda, E. Katragkou, E. Giannakaki, R. E. Mamouri, E. Gerasopoulos, and C. Zerefos. (2007) Aerosol Lidar observations and model calculations of the Planetary Boundary Layer evolution over Greece, during the March 2006 Total Solar Eclipse. Atmos. Chem. Phys. 7, 24: 6181-6189.

Anderson R. C., Keefer D. R., and O. E. Myers (1972), Atmospheric Pressure and Temperature Changes During the 7 March 1970 Solar Eclipse, J. Atmos. Sci. 29, 3, 583-587

Anderson, J. (1987), Two eclipses visible from a geostationary weather satellite, J. Roy. Ast. Soc. Canada (ISSN 0035-872X), 81, 83-94.

Anon (1834), Meteorological Observations during the Solar Eclipse of 30th November. 1834, at Boston, Mass, The Boston Medical and Surgical Journal, Dec 10, 1834 American Periodicals Series II, Boston.

Anon (1990). Effect of the annular eclipse on the solar radiation and surface meteorological elements, Chinese J. Geophys., 33, 4, 399-407 (in Chinese with English abstract)
*Phil. Trans. R. Soc. A.*

*Phil. Trans. R. Soc. A.*

# Tables

| Date | Type | Approximate path over land | Quantities measured | Approach | References |
|---|---|---|---|---|---|
| 1834 Nov 30 | T | USA and Canada | p, T | Boston, Massachusetts | Anon (1834) |
| 1836 May 15 | A | Central America, Caribbean, N and E Europe | u, clouds | Qualitative observations at Greenwich, UK | Birt (1836) |
| 1887 Aug 19 | T | E Europe, Russia, China, Japan | P, T, RH, u, clouds | Multi site | Hesehus and Schönrock in Anderson et al (1972) |
| | | | | Multi instrumented site | Upton and Rotch (1887) |
| 1889 Jan 1 | T | USA and Canada | P, T, u, solar radiation | Multi sites | Upton and Rotch (1893) |
| 1900 May 28 | T | C America, S USA, Spain, N Africa | p, T, u ?? | Multi sites | Clayton (1901) |
| 1912 October 10 | T | S America | AE | Ionisation, PG, conductivity, ion concentration | Knoche and Laub (1917) |
| 1914 August 21 | T | Scandinavia, E Europe, Black Sea region | P, T, u, RH, AE, clouds | Multi sites at magnetic observatories worldwide | Bauer and Fisk (1916) |
| 1916 February 3 | T | N tip of S America, Caribbean | T, photometric measurements | Falcón state, Venezuela | Penaloza-Murillo (2002) |
| 1918 June 8 | T | USA | Clouds, solar fluxes, RH, p, u | Multi sites | Fergusson (1919); Kimball and Fergusson (1919) |
| 1919 May 29 | T | S America, Africa | T, RH, p, magnetic effects, AE | Multi sites | Bauer (1919) |
| | | | | Bipolar conductivity and PG in Brazil | Mauchly and Thomson (1920) |
| 1921 April 8 | A | N UK, N Norway | Solar fluxes. UV, T, RH, p, clouds, polarisation | Davos, Switzerland | Dorno (1925) |
| | | | | Multi site | Billham (1921) |
| 1925 January 24 | T | N USA (Great Lakes) | Solar fluxes, T, u *Aloft*: T, RH, u | Washington DC | Kimball (1925) |
| | | | | Multi sites | Varney (1925) |

| | | | | | Kite flights from two sites | Samuels (1925) |
|---|---|---|---|---|---|---|
| 1936 June 19 | | T | Greece, Russia and China | AE | Russia | Litvinov, in Aplin and Harrison (2003) |
| 1952 February 25 | | T | C Africa, Gulf states, Russia | Solar fluxes with filters | | Abbott (1952) |
| 1954 June 30 | | T | N USA, Canada, Iceland, S Scandinavia, Russia, C Asia | T | Kew Observatory | Harrison (2016) |
| | | | | | Multi sites in UK and Sweden | Morris Bower (1955); Botley (1955) |
| 1955 December 14 | | T | Sri Lanka, SE Asia | Solar fluxes | Poona, India | Jagannathan et al in Aplin and Harrison (2003) |
| 1963 July 20 | | T | Japan, Alaska and Canada | Radiative fluxes | | Pruitt et al (1963) |
| 1970 March 7 | | T | C America, USA E coast and Canada | *Surface:* T, IR, p  *Aloft*: u, T | Rockets; Wallops Island, USA | Quiroz and Henry (1973); Staffanson et al (1973) |
| | | | | | IR thermometer | Hall (1980) |
| | | | | | Lee, Florida | Anderson et al (1972) |
| 1972 July 10 | | T | E Russia, Alaska, Canada | T | Canada | Finkelstein (1973) |
| | | | | | Northern Canada | Stewart and Rouse (1974) |
| 1973 June 30 | | T | NE S America, C Africa | p, T, u | Lee, Florida | Anderson and Keefer (1973) |
| 1976 October 23 | | T | Tanzania, Australia | P, T, u, turbulent fluxes | Gravity wave detection at multiple Australian sites | Goodwin and Hobson (1978) |
| | | | | | New South Wales, Australia | Antonia et al (1979) |
| 1980 February 16 | | T | S Africa, India and China | *Surface*: p, T, RH, fluxes, $O_3$, AE  *Aloft*: T, $O_3$ | Array of four sensors ~1000m apart for gravity wave detection near Saskatchewan, Canada | McIntosh and Revelle (1984) |
| | | | | | multiple sites | Verma et al (1980) |

| | | | | Potential gradient and air conductivity at Raichur, India | Manohar et al (1995) |
|---|---|---|---|---|---|
| 1981 July 31 | T | Kazahkstan and Russia | Fluxes, T | | Kidd (1985) |
| 1983 June 11 | T | SE Asia | P | Gravity waves; multiple sites | Seykora et al (1985) |
| 1984 May 30 | A | Mexico and Sahara Desert | T, RH, u, evaporation, solar radiation, net radiation, radar reflectivity; | Oklahoma, USA | Rabin and Doviak (1989) Trapasso and Kinkel (1984) |
| 1986 October 3 | T | Sea between Greenland and Iceland. Partial over N America | Infrared and visible radiation | Satellite observations | Anderson (1987) |
| 1987 September 23 | A | C and E Asia, S Pacific | p, T, u, RH, radiative fluxes, $O_3$ | China | Anon (1990) |
| 1991 July 11 | T | C America | *Surface:* p, T, u, RH profiles, UV, radiative fluxes, $O_3$ and $CO_2$ fluxes, columnar $O_3$, aerosol, soil T and heat fluxes, ground heat flux *Aloft:* T, u | Sonic anemometry, aircraft, instrumented tower at cotton field (Fresno, California) | Mauder et al (2007) |
| | | | | Multi site; effect of land use on response | Brazel et al (1993) |
| | | | | Multi site; one radiosonde launch | Fernandez et al (1993) |
| 1994 May 10 | A | C America, USA, Morocco | Radiative fluxes, T, radar, sodar, | Multi site | Segal et al (1996) |
| | | | | Micrometeorological | Eaton et al (1997) |

| | | | | scintillometers, u | measurements and optical properties; White Sands Missile Range, New Mexico | |
|---|---|---|---|---|---|---|
| 1994 November 3 | T | | S America | Sg, T, RH, u | Coronel Oviedo, Paraguay | Fernandez et al (1996) |
| 1995 October 24 | T | | India and SE Asia | Multiwavelength solar radiation (visible and UV), T, RH, u speed, IR spectra | Multi site; aerosol optical properties derived | Niranjan and Thularsiraman (1998) |
| | | | | | Multi site; estimate water vapour and $O_3$ | Jain et al (1997) |
| 1998 February 26 | T | | | Photometric measurements | Atmospheric optical properties; Falcón state, Venezuela | Penaloza-Murillo (2002) |
| 1999 August 11 | T | | NW and SE Europe, Iran and India | *Surface:* p, T, u, RH, UV, AE, radiative fluxes, $O_3$ and $CO_2$ fluxes, columnar $O_3$, aerosol, soil T and heat fluxes, ground heat flux *Aloft:* T, u *Upper atmosphere*: GPS, TEC(?), ionosondes | Akola, India; Belgrade, Serbia; Rothamstead, UK; Szczawnica, Poland | Dolas et al (2002); Kolarz et al (2005); Leeds-Harrison et al (2000); Szalowski (2002) |
| | | | | | Multi site | Ahrens et al (2001); Aplin and Harrison (2003); Hanna (2000); Kalafotoglu et al (2007); Winkler et al (2001) |
| | | | | | Lidar | Kolev et al (2005); Satyanarayana et al (2002) |
| | | | | | Micrometeorological measurements (sonic anemometry and/or turbulent fluxes) | Anfossi et al (2004); Aplin and Harrison (2003); Foken et al (2001) |
| | | | | | Ionosondes and high frequency radar | Altadill et al (2001); Farges et al (2001) |
| | | | | | GPS effects | Filjar (2001) |
| | | | | | Radiosonde launches over Trivandrum, India | Satyanarayana et al (2002) |

| | | | | Microbarographs for gravity wave detection over France | Farges et al (2003) |
|---|---|---|---|---|---|
| 2001 June 21 | T | Southern Africa, Madagascar | T, RH, u, global solar radiation, soil heat flux, soil temperatures | Multi site in Zimbabwe | Milford (2001) and this paper |
| 2002 December 4 | T | Angola and Australia | | Modelling stratospheric UV, gravity waves, surface T changes | Eckermann et al (2007) |
| 2003 May 31 | A | Greenland, Iceland and N UK | p | Gravity waves at Inverness, UK | Marlton et al (2016) |
| 2003 November 23 | T | Antarctica | p, T, u, RH, radiative fluxes, snow temperatures | Effect on radiation balance at East Dronning Maud Land, Antarctica | Kameda et al (2009) |
| 2006 March 29 | T | Brazil, N Africa, S Europe and the Black Sea states | *Surface:* p, T, u, RH, radiative fluxes, rain, soil T and moisture, solar spectral irradiance, UV, lidar, $O_3$, NOX *Upper atmosphere:* TEC, peak electron density height | Manavgat, Turkey; Ibadan, Nigeria | Nymphas et al (2009), Uddin et al (2007) |
| | | | | Upper atmosphere effects; search for gravity waves | Zerefos et al (2007) |
| | | | | Spectral effect of limb darkening | Kazadzis et al (2007) |
| | | | | Sonic anemometer; multiple sites | Founda et al (2007) |
| | | | | UV and photosynthetically active radiation; single site | Kazantzidis et al (2007); Blumthaler et al (2006) |
| | | | | Aerosol lidar for boundary layer | Amiridis et al (2007) |

| | | | | investigation | |
| --- | --- | --- | --- | --- | --- |
| | | | | Surface ozone and $NO_2$ | Tzanis et al (2007) |
| 2008 August 1 | T | Canada, Greenland, Russia and China | T, p, u, RH, AE, net radiation | Micrometeorological measurements; multiple sites on Svalbard including on a 30m tower | Sjöblom (2010) |
| | | | | Microbarographs for gravity waves; multiple sites | Marty et al (2013) |
| | | | | Atmospheric potential gradient, sferics; Calcutta, India | De et al (2010) |
| 2009 July 22 | T | India, SE Asia, China | NOX, $O_3$, T | Air pollution monitoring stations in Korea | Kwak et al (2011); Jeon et al (2011) |
| 2010 January 10 | T | Easter Island and S tip of S America | *Surface*: T, p, U, RH, rain, fluxes, Doppler sodar, $O_3$, NOX, turbulent fluxes *Aloft*: T, p, u, RH winds and $O_3$ 20-70km *Upper atm*: TEC (derived from GPS), | Tamil Nadu, India; Cochin, India | Bhat and Jagganthan (2012); Jayakrishnan et al (2013) |
| | | | | Radiosonde launches; multiple sites | Muraleedharan et al (2011) |
| | | | | Near-surface ozone and NOx; multiple sites | Girach et al (2012) |
| | | | | Multi site ionospheric study | Kumar et al (2011) |
| | | | | Sea breeze circulation; turbulent fluxes; rocket, radiosonde and surface measurements on day of eclipse and preceding day; Thumba, India | Subrahamanyan et al (2011); KiranKumar et al (2013); Namboodiri et al (2011) |

| | | | | Gamma and cosmic rays with temperature and pressure at Rameswaram, India | Bhaskar et al (2011) |
| | | | | Radiosonde launches; Doppler sodar; 15m tower; Gadanki, India | Ratnam et al (2010) |

*Table 1 Meteorological observations during solar eclipses, sorted by date of eclipse. All total (t) or annular (a) eclipses have a partial region for several thousand km around them Contemporary country and region names are used. Modelling studies are not included except when there are no observations available. Where no detail is given in the Approach column it indicates that "standard" meteorological measurements only were made. Abbreviations. p: atmospheric pressure, T: temperature, u: wind speed and direction, RH: relative humidity or other water vapour, UV: ultraviolet radiation, AE: atmospheric electricity, GPS: global positioning system, TEC: total electron content, NOX, nitrogen oxides*

| Type of model | Eclipse the model was applied to, and comments | Reference |
|---|---|---|
| NOGAPS-ALPHA: US Department of Defense, high altitude, radiative scheme | 2002; referenced to model with no eclipse; predicted gravity wave, temperature and ozone changes; no observations available | Eckermann et al (2007) |
| UK Met Office, high temporal and spatial resolution | 1999; referenced to model with no eclipse; saw circulation changes consistent with observations | Gray and Harrison (2012) |
| German Weather Service (hydrostatic model) | 1999; referenced to 11 August 1998; pan-European effects | Gross and Henze (1999) |
| BLM regional weather prediction model (German Military Geophysical Office) | 1999; low spatial resolution (60km grid square); referenced to model with no eclipse | Prenosil (2000) |
| KAMM non hydrostatic mesoscale model | 1999; referenced to clear sky model; local effects | Vogel et al (2001) |
| Monte Carlo photon scattering model to predict changes in the visible radiation spectrum | 1978 (simple model) 2006 (more complex 3D model) | Shaw (1978) Emde and Mayer (2007) |
| Meteorological radiation model (MRM) | 2006; predictions of global and diffuse solar radiation compared satisfactorily to measurements | Psiloglou and Kambezedis (2007) |
| Limb darkening model | 1999; spectral irradiance, global and diffuse solar radiation compared satisfactorily to measurements | Koepke et al (2001) |

*Table 2 Summary of eclipse modelling work*

| Year of eclipse | Instrumentation | Results | Reference |
|---|---|---|---|
| 1970 | High-resolution barometer in totality | Main periodicity (Fourier analysis) was 89 min, 25 Pa | Anderson et al (1972) |
| 1976 | 4 microbarograph array, separation 100-1000 km | 22-24 min periodicity decaying to 40 min – but waves had started before eclipse (Zirker, 1991) | Goodwin and Hobson (1978) |
| 1999 | High-resolution barometer in totality | 35 min periodicity (Fourier analysis) decaying to longer period and smaller amplitude. Signals present with 99% confidence compared to non-eclipse times | Aplin and Harrison (2003) |
| 1979 | 4 infrasound detectors, separation ~ 1000m | 12 Pa signal with 2 min periodicity. Not thought to be gravity waves. | Mcintosh and Revelle (1984) |
| 1999 | 3 microbarograph array, separation ~100km | 5 Pa signal with 9-12 min periodicity | Farges et al (2003) |
| 1983 | 4 microbarograph array, one instrument in totality, separation ~1000km | 10–15 Pa fluctuation of period 4 hours | Seykora et al (1985) |
| 2006 | Analysed temperature and relative humidity fluctuations | No effects seen | Zerefos et al (2007) |

*Table 3 Summary of ground-based measurements of eclipse-induced gravity waves from the troposphere and stratosphere*